\documentclass[twocolumn,twoside,slac_two,floatfix]{revtex4}
\usepackage{graphicx}
\usepackage{fancyhdr}
\RequirePackage{xspace}
\usepackage{relsize}
\def\babar{\mbox{\slshape B\kern-0.1em{\smaller A}\kern-0.1em
    B\kern-0.1em{\smaller A\kern-0.2em R}}}
\def\CP                {\ensuremath{C\!P}\xspace}
\def\piz   {\ensuremath{\pi^0}\xspace}

\def\Kstarp  {\ensuremath{K^{*+}}\xspace}
\def\Dbar    {\kern 0.2em\overline{\kern -0.2em D}{}\xspace}
\def\Dz      {\ensuremath{D^0}\xspace}
\def\Dzb     {\ensuremath{\Dbar^0}\xspace}
\def\Dstarp  {\ensuremath{D^{*+}}\xspace}

\def\Y#1S{\ensuremath{\Upsilon{(#1S)}}\xspace}

\def\invfb   {\ensuremath{\mbox{\,fb}^{-1}}\xspace}
\newcommand{\mev}{\ensuremath{\mathrm{\,Me\kern -0.1em V}}\xspace}
\newcommand{\mevc}{\ensuremath{{\mathrm{\,Me\kern -0.1em V\!/}c}}\xspace}
\newcommand{\mevcc}{\ensuremath{{\mathrm{\,Me\kern -0.1em V\!/}c^2}}\xspace}
\newcommand{\gevc}{\ensuremath{{\mathrm{\,Ge\kern -0.1em V\!/}c}}\xspace}
\newcommand{\gevcc}{\ensuremath{{\mathrm{\,Ge\kern -0.1em V\!/}c^2}}\xspace}

\def\pep2{PEP-II}

\newcommand{\rsdecay}{\mbox{\ensuremath{\Dz \to K^{-} \pi^{+} \pi^{0}}}}
\newcommand{\wsdecay}{\mbox{\ensuremath{\Dz \to K^{+} \pi^{-} \pi^{0}}}}
\newcommand{\wsdecayb}{\mbox{\ensuremath{\Dzb \to K^{-} \pi^{+} \pi^{0}}}}
\newcommand{\dm}{\ensuremath{\Delta m}}
\newcommand{\mKpp}{\ensuremath{m_{K\pi\pi^{0}}}}
\newcommand{\tKpp}{\ensuremath{t_{K\pi\pi^{0}}}}
\newcommand{\rightsign}{RS}
\newcommand{\wrongsign}{WS}

\begin{document}

\title{\boldmath \Dz-\Dzb\ mixing results from \babar\
by analysis\\ of \wsdecay\ Dalitz-plot regions}

\author{Michael G. Wilson, for the \babar\ Collaboration}
\affiliation{Univ. of California at Santa Cruz, Institute for Particle Physics, Santa Cruz, California 95064, USA}

\begin{abstract}
We present a preliminary search for \Dz-\Dzb\ mixing using the decays \wsdecay,
additionally presenting Dalitz-plot distributions and a measurement of the
branching ratio for this mode.  A new tagging technique is used to produce
the doubly Cabibbo-suppressed Dalitz plot, which in turn is used to motivate the method used
for the $D$-mixing search.  We analyze 230.4\invfb of data collected from the \babar\
detector at the \pep2 collider.
Assuming \CP\ conservation, we find $R_M < 0.054\%$ with 95\% confidence,
and we estimate that the data are consistent with no mixing at a 4.5\% confidence level.
We present $D$-mixing results both with and without the assumption of \CP\ conservation.
\end{abstract}

\maketitle

\thispagestyle{fancy}


\section{Introduction}
Although $K$ and $B$ mixing are well established, $D$ mixing
has yet to be observed.  As this particular mixing phenomenon
is sensitive to new physics in a complementary manner to
the $K$ and $B$ systems, it is an essential test of the
completeness of the Standard Model.  We present
preliminary results of a new mixing analysis using the decays
\wsdecay, taking into account the resonant structure of the doubly
Cabibbo-suppressed contributions.  This is the first
search for mixing using either this decay mode or this
technique.

G.~Burdman and I.~Shipsey have written
a thorough review of $D$-mixing predictions~\cite{Burdman:2003rs};
accurate predictions are difficult to obtain
because of significant contributions from long-distance effects.

Using 230.4\invfb of data collected from the \babar\
detector at the \pep2 collider, corresponding to
approximately 300 million $c\bar{c}$ events,
we obtain a pure sample of \Dz\ candidates
by reconstructing the decays
\begin{eqnarray}
\Dstarp \to & \Dz\pi^{+}_{s} & \nonumber \\
            & \Dz \to & K^{\mp}\pi^{\pm}\pi^{0}\; \; \; \; \; \textrm{(+ C.C.).}
\label{eq:decaychain}
\end{eqnarray}
The flavor of the \Dz\ candidate at production is carried by the
charge of the associated $\pi^{\pm}_{s}$.  The large
sample of Cabibbo-favored (CF) decays \rsdecay\ is used
both as a means for reducing systematic uncertainties in the probability
density functions (PDFs) used to describe the
\textit{wrong-sign} (WS) decays \wsdecay\ and as a normalization
mode when determining the WS decay rate.

We separate correctly reconstructed decays from background,
and distinguish doubly Cabibbo-suppressed (DCS)
contributions from CF mixed contributions, by means of an
(unbinned) extended maximum likelihood fit.
PDFs are fit to the three distributions
$\{\mKpp,\dm,\tKpp\}$, where \mKpp\ is the invariant mass
of the \Dz\ candidate, \dm\ is the invariant mass difference
between the \Dstarp\ and \Dz\ candidates,
and \tKpp\ is the candidate decay time.

\section{\boldmath \wsdecay\ Branching Ratio}

In a nonleptonic search for $D$ mixing, DCS contributions
obscure signs of mixing in the final state.  To the
extent that the DCS rate is low for a particular mode
compared to the corresponding CF rate, there is greater
sensitivity to a potential mixing signal.
The branching ratio
\begin{equation}
R = \frac{\Gamma(\wsdecay)}{\Gamma(\rsdecay)}
\end{equation}
is measured using a maximum likelihood fit to the
distributions $\{\mKpp,\dm\}$, and we find the preliminary
result
\begin{equation}
R = (0.214 \pm 0.008\,(\textrm{stat}) \pm 0.008\,(\textrm{syst})) \%\textrm{.}
\label{eq:br}
\end{equation}
This result is consistent with that reported by the
Belle Collaboration last year~\cite{Tian:2005ik}.
Comparing this to the corresponding branching ratio
for the decay $\Dz\to K^+\pi^-$~\cite{Eidelman:2004wy},
\begin{equation}
R(K\pi) = (0.362 \pm 0.029) \%\textrm{,}
\end{equation}
we find that an analysis of
\wsdecay\ may in fact have more sensitivity to mixing
than the standard analysis of the decays $\Dz\to K^+\pi^-$.
We note that while the branching ratio in
Eq.~\ref{eq:br} may contain contributions from
$D$ mixing (known to be small), the level of these contributions cannot be
determined without an analysis of the WS decay-time distribution.

\section{Event-Level Tagging and DCS Resonance Contributions}

\begin{figure*}[p!]
\begin{center}
 \begin{tabular}{p{0.5\linewidth}p{0.5\linewidth}}
  \begin{center}
   \includegraphics[width=80mm]{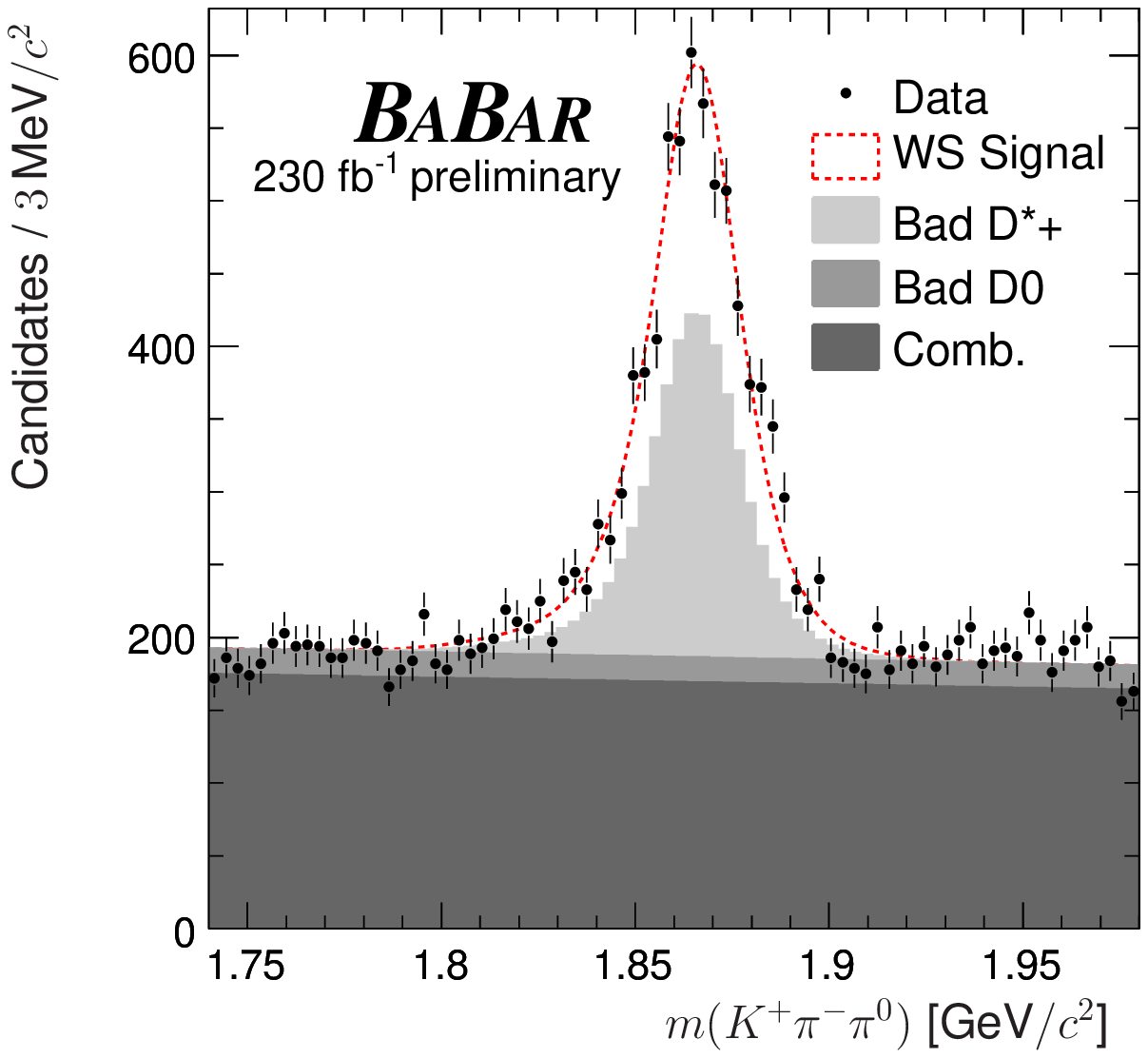}
  \end{center}
  &
  \begin{center}
   \includegraphics[width=80mm]{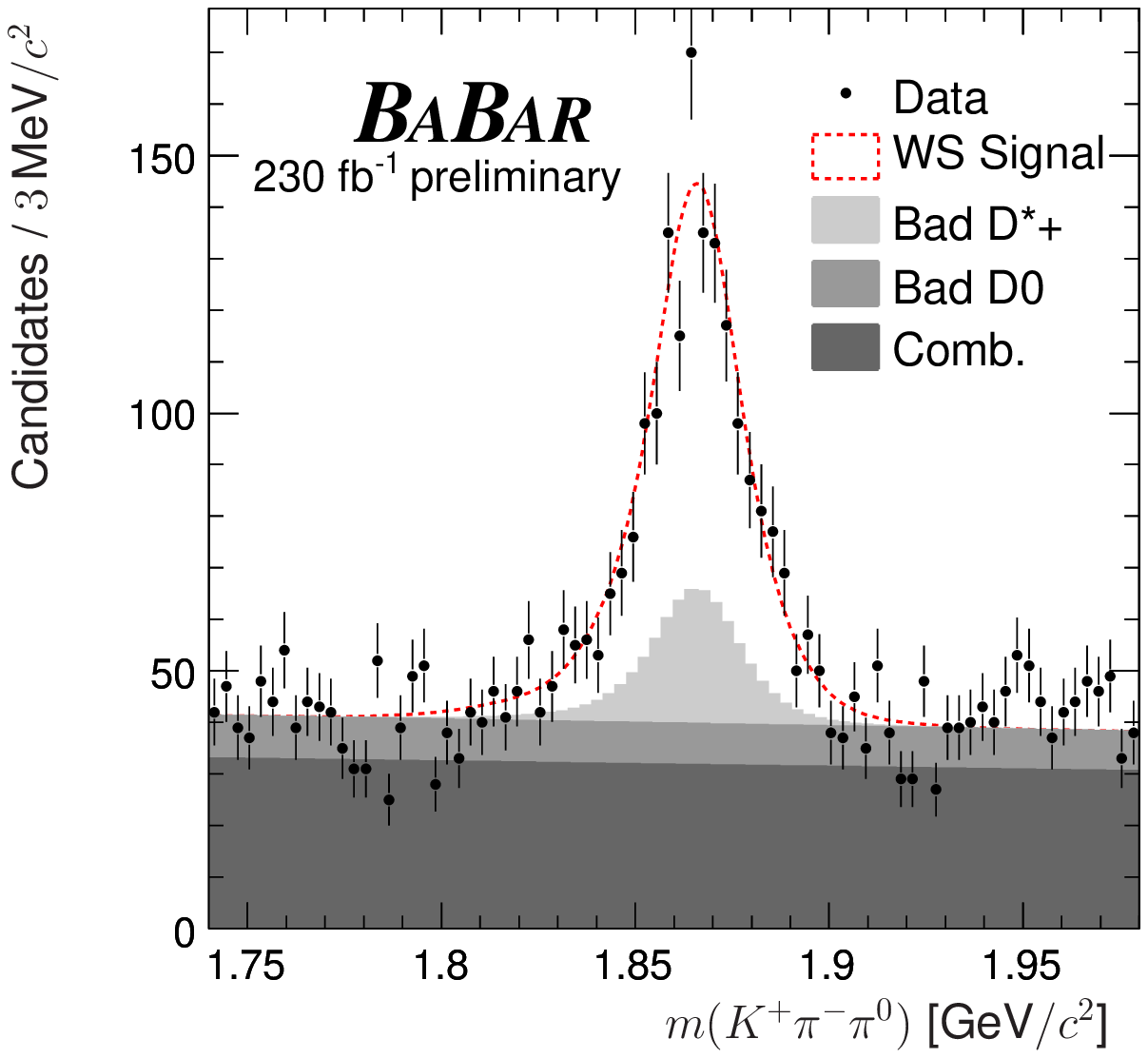}
  \end{center}
 \end{tabular}
\caption{
The \mKpp\ distribution with fitted PDFs
overlaid for the WS data sample used to search for $D$
mixing (left) and for the sample used to obtain the
Dalitz-plot distributions (right).  The sample on
the right requires a second, event-level, flavor tag.
Both distributions require
$0.1444 < \dm < 0.1464\gevcc$.}
\label{fig:mdz}
\end{center}
\end{figure*}

\begin{figure*}[p!]
\begin{center}
 \begin{tabular}{p{0.5\linewidth}p{0.5\linewidth}}
  \begin{center}
   \includegraphics[width=80mm]{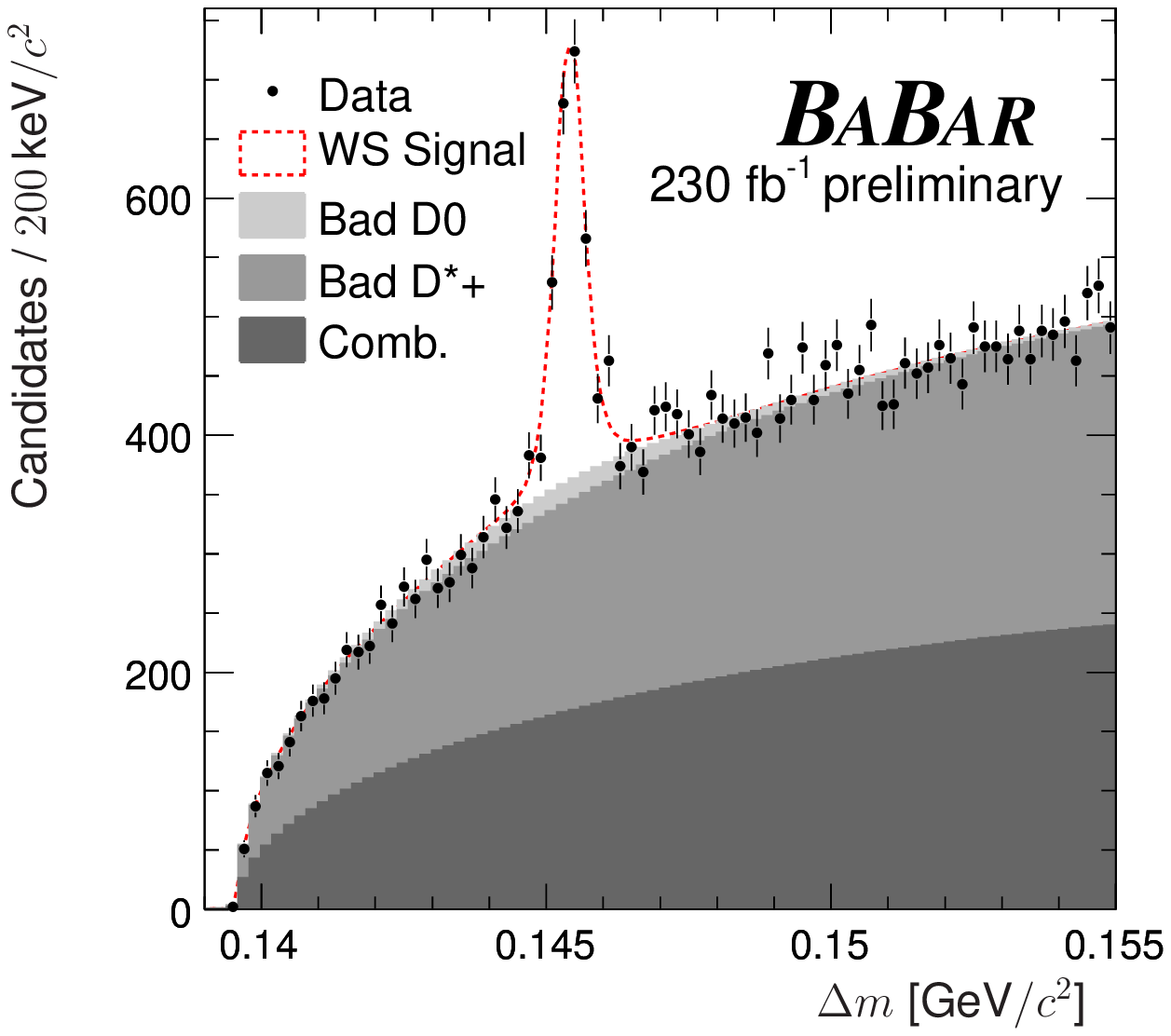}
  \end{center}
  &
  \begin{center}
   \includegraphics[width=80mm]{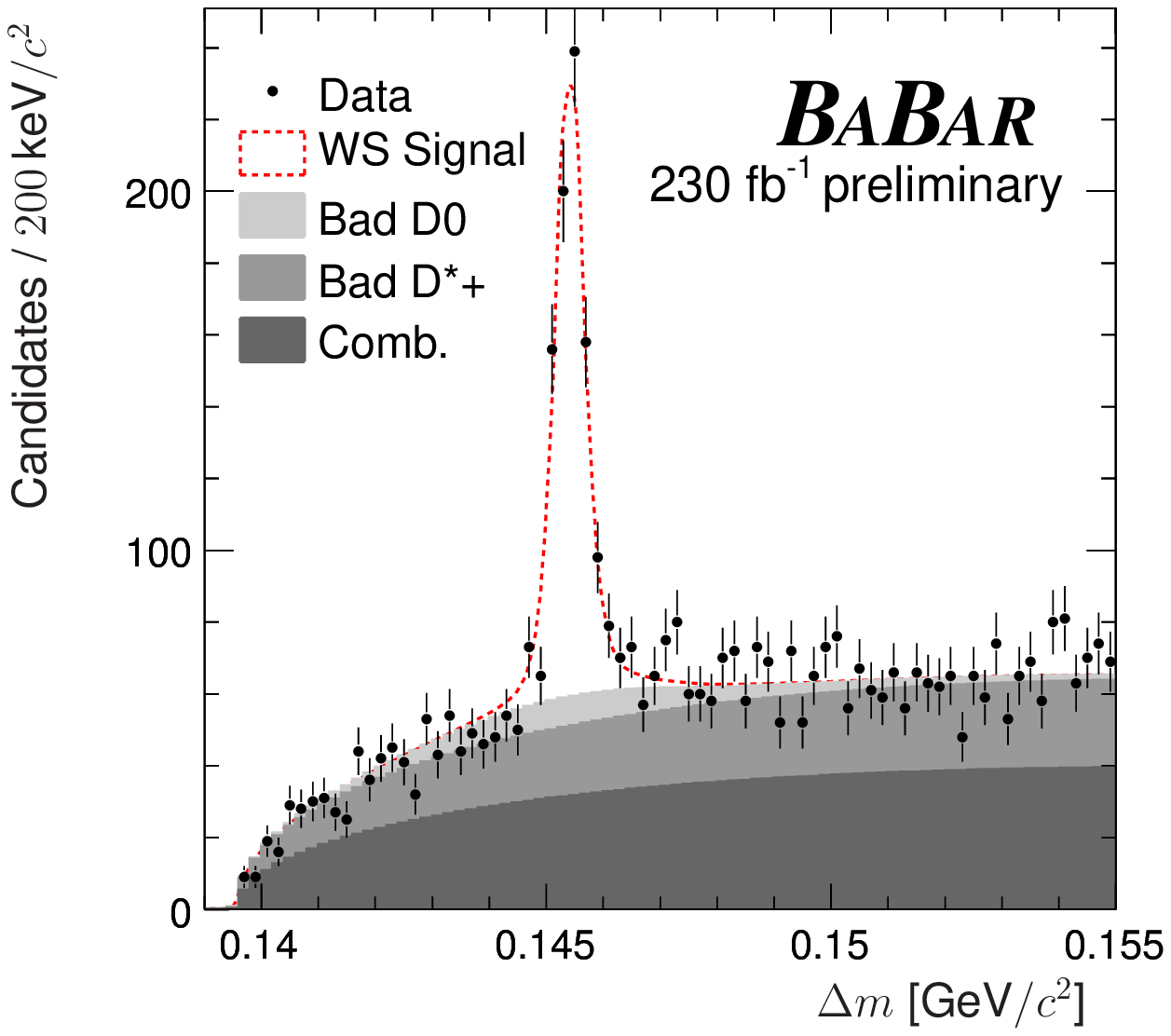}
  \end{center}
 \end{tabular}
\caption{
The \dm\ distribution with fitted PDFs
overlaid for the WS data sample used to search for $D$
mixing (left) and for the sample used to obtain the
Dalitz-plot distributions (right).  The sample on
the right requires a second, event-level, flavor tag.
Both distributions require
$1.85 < \mKpp < 1.88\gevcc$.}
\label{fig:deltam}
\end{center}
\end{figure*}

\begin{figure*}[t!]
\begin{center}
 \begin{tabular}{p{0.5\linewidth}p{0.5\linewidth}}
  \begin{center}
   \includegraphics[width=80mm]{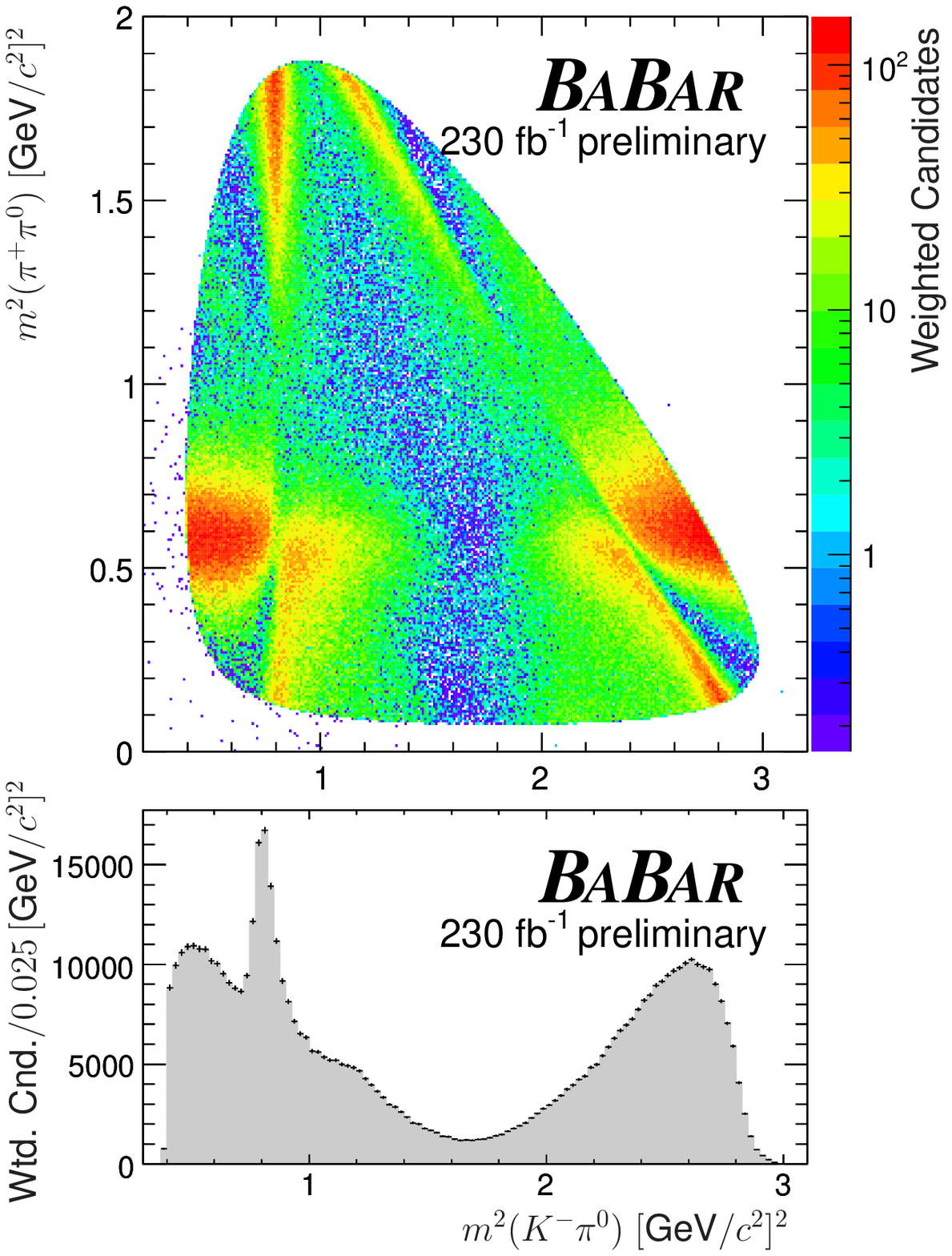}
  \end{center}
  &
  \begin{center}
   \includegraphics[width=80mm]{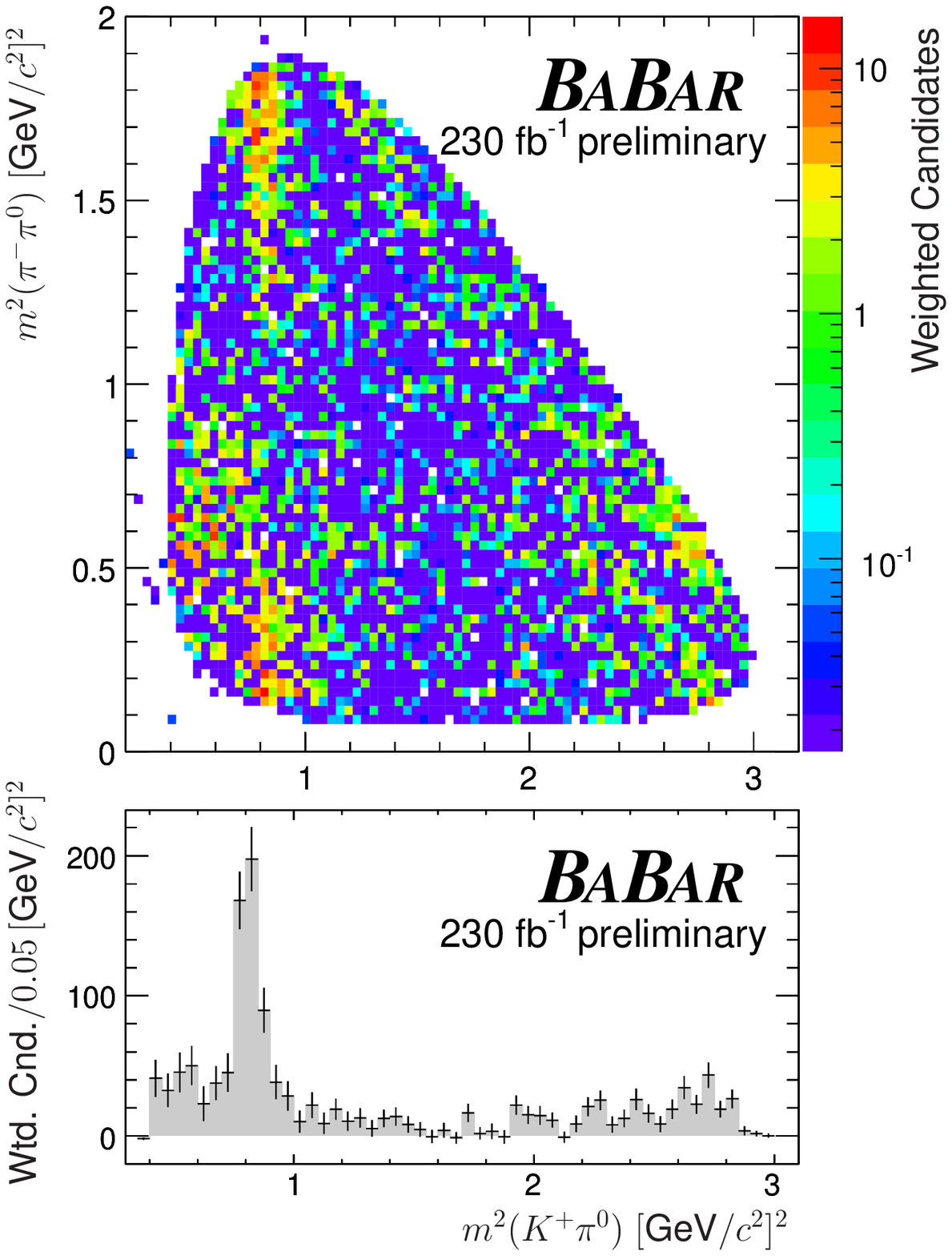}
  \end{center}
 \end{tabular}
\caption{
Dalitz plot and one-dimensional projection onto $m^2_{K\piz}$ for
the CF decay (left) and the DCS decay (right).  A statistical
background subtraction has been applied (\textit{i.e.} candidates have been weighted),
and distributions
have been corrected for efficiency variations as a function of the Dalitz
plot.  The shaded histograms in the lower plots are projections of
the Dalitz plots above them, with error bars superimposed.
Candidates have been selected requiring the event-level tag.}
\label{fig:m2Kpiz}
\end{center}
\end{figure*}

Just as an analysis of the decay \wsdecay\
may offer greater sensitivity to a mixing
signal than that of $\Dz\to K^+\pi^-$
because of the lower relative DCS contribution,
regions of the decay phase space
(\textit{i.e.}, the Dalitz plot) may have greater
sensitivity to mixing to the extent that the
DCS contribution in those regions is relatively low.
However, obtaining an accurate distribution of events
in the Dalitz plot is challenging because of the
large number of background events still remaining after
requiring event-selection criteria.  In particular,
a large component of the background is from
real \Dz\ decays having a misassociated $\pi^{\pm}_{s}$.
Since this background is from real \rsdecay\ decays, it populates
the Dalitz plot with structure that tends to obscure that
of the signal.  This peaking background is marked ``Bad D*+''
in Figs.~\ref{fig:mdz} and~\ref{fig:deltam}.

To suppress peaking background from \rsdecay\ decays, we use
an event-level tag in addition to the primary $\pi^{\pm}_{s}$
flavor tag.  This tag is determined by searching
the event hemisphere opposite that in which the
\Dstarp\ candidate is reconstructed for any of the following:
$K^\pm$, $\pi^{\mp}_{s}$, $e^{\mp}$, or
$\mu^{\mp}$.  If the charge of one of these candidate is consistent
with the hypothesis of hadronization and decay of a
$\bar{c}$~quark in that event hemisphere, then the event is tagged.
The signal efficiency after applying this second, opposite-side, tag
is 46.4\%, while the peaking-background efficiency is 10.9\%.
The effect of requiring this tag is shown in
Figs.~\ref{fig:mdz} and~\ref{fig:deltam}.

After requiring the opposite-side tag, correcting
for efficiency variations as a function of the Dalitz
plot, and performing a
statistical background subtraction~\cite{Pivk:2004ty}
based on a maximum likelihood fit to 
$\{\mKpp,\dm\}$, we obtain the Dalitz plots in
Fig.~\ref{fig:m2Kpiz}.  The CF Dalitz plot
is qualitatively different from the WS Dalitz plot, which
is assumed to contain primarily
DCS contributions\footnote{We assume that possible contributions
from $D$ mixing cannot be distinguished by eye in the Dalitz plots
shown.}.  While CF decays proceed primarily through the resonance
$\Dz\to K^-\rho^+$, DCS decays proceed primarily through the resonance
$\Dz\to \Kstarp\pi^-$.  We use this observation to maximize sensitivity
to a potential $D$-mixing signal as described in the next Section.

\section{Preliminary Results of a $D$-mixing Search}

\begin{figure}
\begin{center}
\includegraphics[width=80mm]{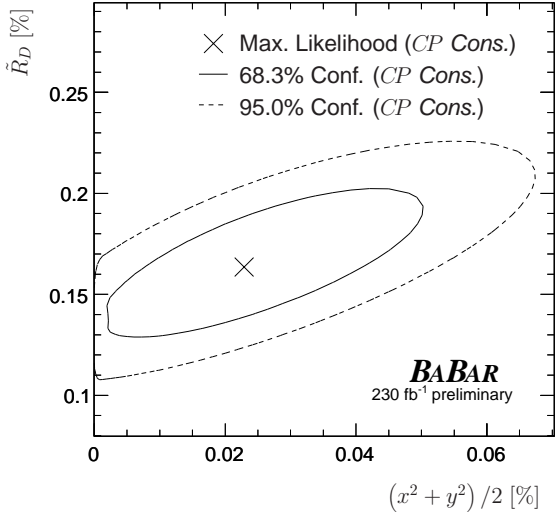}
\caption{
Contours of constant $\Delta\log\mathcal{L}=1.15,3.0$ in terms of
the doubly Cabibbo-suppressed rate and the integrated
mixing rate.  The upward slope of the contour indicates
negative interference.}
\label{fig:rmixrd}
\end{center}
\end{figure}

The two mass eigenstates
\begin{equation}
\label{eq:massstates}
 |D_{1,2}\rangle = p|\Dz\rangle \pm q|\Dzb\rangle
\end{equation}
generated by mixing dynamics have different
masses $(m_{1,2})$ and widths $(\Gamma_{1,2})$, and we
parameterize the mixing process with the quantities
\begin{equation}
x \equiv 2\frac{m_{2} - m_{1}}{\Gamma_{2} + \Gamma_{1}},\ \ \ \ \ %
y \equiv \frac{\Gamma_{2} - \Gamma_{1}}{\Gamma_{2} + \Gamma_{1}}%
\textrm{.}
\end{equation}
If \CP\ is not violated, then $|p/q|=1$.
For a multibody \wrongsign\ decay, the time-dependent decay
rate, relative to a corresponding \textit{right-sign} (RS) rate, is
well approximated by
\begin{eqnarray}
\label{eq:tdratemult}
 & {\displaystyle\frac{\Gamma_{\textrm{\scriptsize WS}}(t)}{\Gamma_{\textrm{\scriptsize RS}}(t)} =%
   \tilde{R}_D + \alpha\tilde{y}'\sqrt{\tilde{R}_D}\,(\Gamma t)%
   + \frac{\tilde{x}'^2+\tilde{y}'^2}{4}(\Gamma t)^2 } & \\
 &  0 \leq \alpha \leq 1\textrm{,} & \nonumber
\end{eqnarray}
where the tilde indicates quantities that have been integrated over the selected
phase-space regions.  Here, $\tilde{R}_D$ is the integrated DCS branching ratio;
$\tilde{y}' = y\cos\tilde{\delta} - x\sin\tilde{\delta}$ and
$\tilde{x}' = x\cos\tilde{\delta} + y\sin\tilde{\delta}$,
$\tilde{\delta}$ is an unknown integrated strong-phase difference;
$\alpha$ is a suppression factor that accounts for
strong-phase variation over the region; and $\Gamma$ is the
average width.  The
time-integrated mixing rate
$R_M = (\tilde{x}'^2+\tilde{y}'^2)/2 = (x^2+y^2)/2$ is independent of decay mode
and should be consistent among mixing measurements.

\begin{figure}
\begin{center}
\includegraphics[width=80mm]{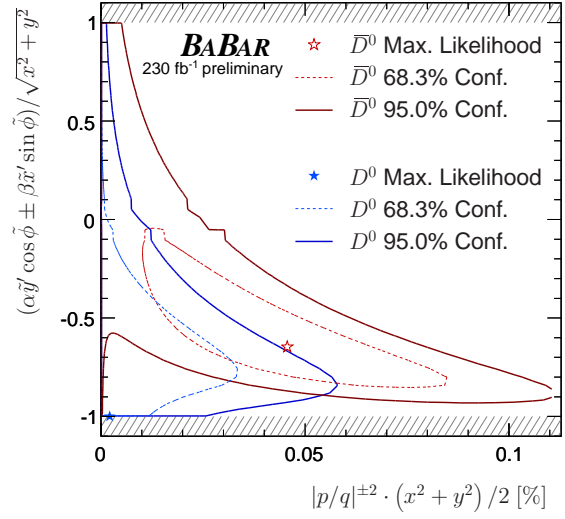}
\caption{
Contours of constant $\Delta\log\mathcal{L}=1.15,3.0$ in terms of
the normalized interference term and the integrated
mixing rate, for the \Dz\ and \Dzb\ samples separately.
The hatched regions are physically forbidden.}
\label{fig:flavorrmixint}
\end{center}
\end{figure}

The necessity of the suppression factor $\alpha$ can be understood as follows.
Suppose the strong-phase difference between DCS decay and mixing is such
that in a particular region of phase space there is positive interference,
while in another region, there is negative interference.  If one
performs a mixing analysis considering the decay-time distributions
of events in both phase-space regions simultaneously, then the combined
decay-time distribution will have a suppressed interference effect
relative to the mixing rate $R_M$, which is nonnegative and invariant
across phase space.

In addition to our search for a \CP-conserving mixing signal, we manifestly
permit
\CP\ violation by fitting to the \wsdecay\ and
\wsdecayb\ samples separately.    We consider \CP\ violation
in the interference between the DCS channel and mixing, parameterized
by an integrated weak-phase $\tilde{\phi}$, as well as \CP\ violation
in mixing, parameterized by $|p/q|$.
We assume \CP\ invariance
in both the DCS and CF rates.  The transformations
\begin{eqnarray}
 & \displaystyle \alpha\tilde{y}' \to \left|\frac{p}{q}\right|^{\pm 1}%
 (\alpha\tilde{y}'\cos\tilde{\phi} \pm \beta\tilde{x}'\sin\tilde{\phi}) & \\
 & \displaystyle (x^2+y^2) \to \left|\frac{p}{q}\right|^{\pm 2}(x^2+y^2) &
\end{eqnarray}
are applied to Eq.~\ref{eq:tdratemult},
using $(+)$ for $\Gamma(\wsdecayb)/\Gamma(\rsdecay)$ and $(-)$ for the
charge-conjugate ratio.  The parameter $\beta$ is analogous to $\alpha$.

Because decays with mixing will
obey the CF-decay resonance structure, and we can qualitatively
see the differences between the CF and DCS Dalitz plots, we
choose regions of the Dalitz plot to analyze in order to maximize
sensitivity to a potential mixing signal.  We do this by
excluding events
with two-body invariant masses in the ranges
$850 < m(K\pi) < 950 \mevcc$ or
$850 < m(K\piz) < 950 \mevcc$.
The signal yields before and after applying
these selection criteria are listed in Table~\ref{tbl:candnumbers}.

\begin{table}[t!]
\caption{
Preliminary signal-candidate yields determined by the
two-dimensional fit to the $\{\mKpp,\dm\}$
distributions
for the \wrongsign\
and \rightsign\ samples.  Yields
are shown for (a) the selected phase-space regions
used in
this analysis and (b) the
entire allowed phase-space region.  Uncertainties
are those calculated from the fit, and no
efficiency corrections have been applied.}
\begin{center}
\begin{tabular}{llrr}
\hline
\multicolumn{2}{c}{\rule{0ex}{3ex}} &
\multicolumn{1}{c}{\Dz\ Cand.} &
\multicolumn{1}{c}{\Dzb\ Cand.} \\
\hline
 \rule{0em}{4.25ex}(\textit{a}) &
 \parbox{2em}{%
 \begin{tabular}{l}
 WS \\
 RS
 \end{tabular}} &
 \parbox{10em}{%
 \begin{tabular}{r}
 $(3.84 \pm 0.36) \times 10^2$ \\
 $(2.518 \pm 0.006) \times 10^5$
 \end{tabular}} &
 \parbox{10em}{%
 \begin{tabular}{r}
 $(3.79 \pm 0.36) \times 10^2$ \\
 $(2.512 \pm 0.006) \times 10^5$
 \end{tabular}} \\
 \rule[-2.75ex]{0em}{7.75ex}(\textit{b}) &
 \parbox{2em}{%
 \begin{tabular}{l}
 WS \\
 RS
 \end{tabular}} &
 \parbox{10em}{%
 \begin{tabular}{r}
 $(7.5 \pm 0.5) \times 10^2$ \\
 $(3.648 \pm 0.007) \times 10^5$
 \end{tabular}} &
 \parbox{10em}{%
 \begin{tabular}{r}
 $(8.1 \pm 0.5) \times 10^2$ \\
 $(3.646 \pm 0.007) \times 10^5$
 \end{tabular}} \\
 \hline
\end{tabular}
\label{tbl:candnumbers}
\end{center}
\end{table}

We search for mixing in the WS decay-time distribution
by constructing a three-dimensional PDF in $\{\mKpp,\dm,\tKpp\}$.
The \wrongsign\ signal PDF in \tKpp\ is
a function based on Eq.~\ref{eq:tdratemult} convolved
with three Gaussians.  We fit the \wrongsign\
PDF to the \tKpp\ distribution allowing yields and
shape parameters to vary.

The results of the decay-time fit, both with and 
without the assumption of \CP\ conservation, are listed in
Table~\ref{tbl:results}.  The statistical uncertainty
of a particular parameter
is obtained by finding its extrema
for $\Delta\log\mathcal{L}=0.5$; in finding
the extrema, the likelihood is kept maximal by
refitting the remaining parameters.
Contours of constant $\Delta\log\mathcal{L}=1.15,3.0$,
enclosing two-dimensional coverage probabilities of
68.3\% and 95.0\%, respectively,
are shown in Figures~\ref{fig:rmixrd}
and~\ref{fig:flavorrmixint}.  The likelihood is maximized
when evaluating $\Delta\log\mathcal{L}$ for a particular
point in two-dimensions.

We note that $\Delta\log\mathcal{L}$ as a function of
the quantity $\textrm{sign}(\alpha\tilde{y}')\times R_M$ is
approximately parabolic.  The two-sided interval
$-0.054\% < \textrm{sign}(\alpha\tilde{y}')\times R_M < 0.054\%$
contains 95\% coverage probability;
thus, we quote $R_M < 0.054\%$
as our upper limit on the integrated mixing rate
under the assumption of \CP\ conservation.

A feature of $\Delta\log\mathcal{L}$ in one dimension
is that it changes behavior near $R_M=0$ because
the interference parameters become unconstrained.  Therefore,
we estimate the consistency of the data with no mixing
using a frequentist method.  Generating 1000 simulated
data sets with no mixing,
each with 58,800 events representing signal
and background in the quantities
$\{\mKpp,\dm,\tKpp\}$, we find 4.5\% of 
simulated data sets have a fitted value of $R_M$
greater than that in the observed data set.
We conclude that the observed data are consistent
with no mixing with 4.5\% confidence.

\begin{table}[ht!]
\caption{
Preliminary mixing results assuming \CP\ conservation
(\Dz\ and \Dzb\ samples are not separated) and
manifestly permitting \CP\ violation (\Dz\ and \Dzb\ samples are
fit separately).  The first listed uncertainty is
statistical, the second is systematic.
$\tilde{R}_D$ is not reported for the latter
case because $\pi_s^{\pm}$
efficiencies were not studied.}
\begin{center}
\begin{tabular}{cr|cr}
\hline
 & \multicolumn{1}{c}{\CP\ conserved} &
 & \multicolumn{1}{c}{\CP violation allowed} \\
\hline
 \rule{0em}{3ex}$R_M$ &
\multicolumn{1}{r}{$(0.023\ \mbox{}^{+0.018}_{-0.014} \pm 0.004) \%$} &
 &
\multicolumn{1}{r}{$(0.010\ \mbox{}^{+0.022}_{-0.007}) \%$} \\
 \rule[-1.25ex]{0em}{1.25ex}$\tilde{R}_D$ &
\multicolumn{1}{r}{$(0.164\ \mbox{}^{+0.026}_{-0.022} \pm 0.012) \% $} &
 &
\multicolumn{1}{r}{} \\
\hline
 \rule{0em}{4.25ex}\rule[-3ex]{0em}{3ex}$\alpha\tilde{y}'$ &
 $-0.012\ \mbox{}^{+0.006}_{-0.008} \pm 0.002$ &
 \parbox{2em}{%
 \begin{tabular}{l}
 $\alpha\tilde{y}'\cos\tilde{\phi}$ \\
 \rule{0em}{2.75ex}$\beta\tilde{x}'\sin\tilde{\phi}$
 \end{tabular}} &
 \parbox{6em}{%
 \begin{tabular}{r}
 $-0.012\ \mbox{}^{+0.006}_{-0.007}$ \\
 \rule{0em}{2.75ex}$0.003\ \mbox{}^{+0.002}_{-0.005}$
 \end{tabular}} \\
\hline
 \rule{0em}{3ex}& &
 \rule[-1.25ex]{0em}{1.25ex}$|p/q|$ & $2.2\ \mbox{}^{+1.9}_{-1.0}$ \\
\hline
\end{tabular}
\label{tbl:results}
\end{center}
\end{table}

\bigskip 
\begin{acknowledgments}
We are grateful for the excellent luminosity and machine conditions
provided by our \pep2\ colleagues, 
and for the substantial dedicated effort from
the computing organizations that support \babar.
The collaborating institutions wish to thank 
SLAC for its support and kind hospitality. 
This work is supported by
DOE
and NSF (USA),
NSERC (Canada),
IHEP (China),
CEA and
CNRS-IN2P3
(France),
BMBF and DFG
(Germany),
INFN (Italy),
FOM (The Netherlands),
NFR (Norway),
MIST (Russia), and
PPARC (United Kingdom). 
Individuals have received support from CONACyT (Mexico), 
Marie Curie EIF (European Union),
the A.~P.~Sloan Foundation, 
the Research Corporation,
and the Alexander von Humboldt Foundation.
\end{acknowledgments}

\bigskip 
\bibliography{references}

\end{document}